%% file: DAC13_EBLOW.tex
\documentclass[9pt]{sig-alternate}
\input{lib/acm_setting}

\graphicspath{{./figs/}}

\begin{document}

\conferenceinfo{DAC'13,} {May 29--June 07 2013, Austin, TX, USA.} 
\CopyrightYear{2013} 
\crdata{978-1-4503-2071-9/13/05}

\title{
E-BLOW: E-Beam Lithography Overlapping aware Stencil Planning for MCC System
}

\iftrue
\numberofauthors{1}
\author{
\alignauthor Bei Yu,\ \  Kun Yuan$^\dag$,\ \  Jhih-Rong Gao,\ \  David Z. Pan\\
\affaddr{
ECE Department, Univ. of Texas at Austin, Austin, TX, USA\\ 
$^\dag$Cadence Design Systems, Inc., San Jose, CA, USA\\
} 
\affaddr{\{bei, jrgao, dpan\}@cerc.utexas.edu}
}
\fi

\maketitle
\thispagestyle{empty}

\input{doc/abstract}

\input{doc/intro}

\input{doc/problem}

\input{doc/1D}

\input{doc/2D}

\input{doc/result}

\input{doc/conclu}

{
\vspace{-.1in}
\scriptsize
\bibliographystyle{IEEEtran}
\bibliography{./Ref/Bei,./Ref/Algorithm,./Ref/EBL,./Ref/EUV,./Ref/Lith,./Ref/MPL,./Ref/Floorplan,./Ref/Partition}
}

\end{document}

%% file: lib/acm_setting.tex
\usepackage{subfigure}
\usepackage{graphicx}
\usepackage{cite}
\usepackage{threeparttable}
\usepackage{ntheorem}
\usepackage{color}
\usepackage{multirow}
\usepackage{amsmath}
\usepackage{comment}

\usepackage{algorithm}
\usepackage{algorithmic}

\theoremstyle{plain}
\theoremheaderfont{\normalfont\bfseries}\theorembodyfont{\slshape}
\theoremsymbol{\ensuremath{\diamondsuit}} \theoremseparator{:}
\newtheorem{mytheorem}{Theorem}

\theoremstyle{plain}
\theoremsymbol{\ensuremath{\clubsuit}}
\theoremseparator{.}
\newtheorem{problem}{Problem}

\theoremsymbol{\ensuremath{\heartsuit}}
\newtheorem{mylemma}{Lemma}

\theoremstyle{change}
\theorembodyfont{\upshape}
\theoremsymbol{\ensuremath{\ast}}
\theoremseparator{}

\theoremheaderfont{\sc}\theorembodyfont{\upshape}
\theoremstyle{nonumberplain} \theoremseparator{}
\theoremsymbol{\rule{1ex}{1ex}}

\theoremstyle{plain} \theoremsymbol{\ensuremath{\clubsuit}}
\theoremseparator{.}

%% file: doc/abstract.tex

\begin{abstract}
Electron beam lithography (EBL) is a promising maskless solution for the technology beyond 14nm logic node.
To overcome its throughput limitation, 
recently the traditional EBL system is extended into MCC system. 
In this paper, we present E-BLOW, a tool to solve the overlapping aware stencil planning (OSP) problems in MCC system.
E-BLOW is integrated with several novel speedup techniques,
i.e., successive relaxation, dynamic programming and KD-Tree based clustering, 
to achieve a good performance in terms of runtime and solution quality.
Experimental results show that, compared with previous works, E-BLOW demonstrates better performance for both conventional EBL system and MCC system.
\end{abstract}


\iftrue
\vspace{-0.05in}
\category{B.7.2}{Hardware, Integrated Circuit} Design Aids
\terms{Algorithms, Design, Performance}
\keywords{Electron Beam Lithography (EBL), Overlapping aware Stencil Planning (OSP), Multi-Column Cell (MCC) System}
\vspace{0.1in}
\fi

%% file: doc/intro.tex
\vspace{-.1in}
\section{Introduction}

As the minimum feature size continues to scale to sub-22nm, the conventional 193nm optical photolithography technology is facing great challenge in manufacturing
\cite{LITH_ICCAD2012_Yu}.
In the near future, double/multiple patterning lithography (DPL /MPL) has become one of viable lithography techniques for 22nm and 14nm logic node
\cite{DPL_ICCAD08_Kahng,TPL_ICCAD2011_Yu,TPL_SPIE2012_Lucas}
.
In the longer future, i.e., for the logic node beyond 14nm, extreme ultra violet (EUV) and electric beam lithography (EBL) are promising candidates for lithographic processes.
However, EUV suffers from the delay due to the tremendous technical barriers such as lack of power sources, resists, and defect-free masks \cite{EUV_SPIE2010_Arisawa}.

EBL system, on the other hand, has been developed for several decades
\cite{EBL_SPIE09_Pfeiffer}.
One of the conventional EBL systems is based on character projection (CP) mode.
Some complex shapes, called \textit{characters}, are prepared on the stencil.
The key idea is that if a pattern is pre-designed on the stencil, it can be printed in one electronic shot,
otherwise it needs to be fractured into a set of rectangles and printed one by one through variable shaped beam (VSB).
Compared with purely VSB mode, in the CP mode the throughput can be improved significantly.
Compared with the traditional lithographic methodologies, EBL has several advantages.
(1) Electron beam can be easily focused into nanometer diameter with charged particle beam,
which can avoid suffering from the diffraction limitation of light.
(2) The price of a photomask set is getting unaffordable.
As a maskless technology, EBL can reduce the manufacturing cost. 
(3) EBL allows a great flexibility for fast turnaround times and even late design modifications to correct or adapt a given chip layout.
Because of all these advantages, EBL is being used in mask making, small volume LSI production, and R\&D to develop the technological nodes ahead of mass production.

\begin{figure}[tb]
  \centering
  \includegraphics[width=0.4\textwidth]{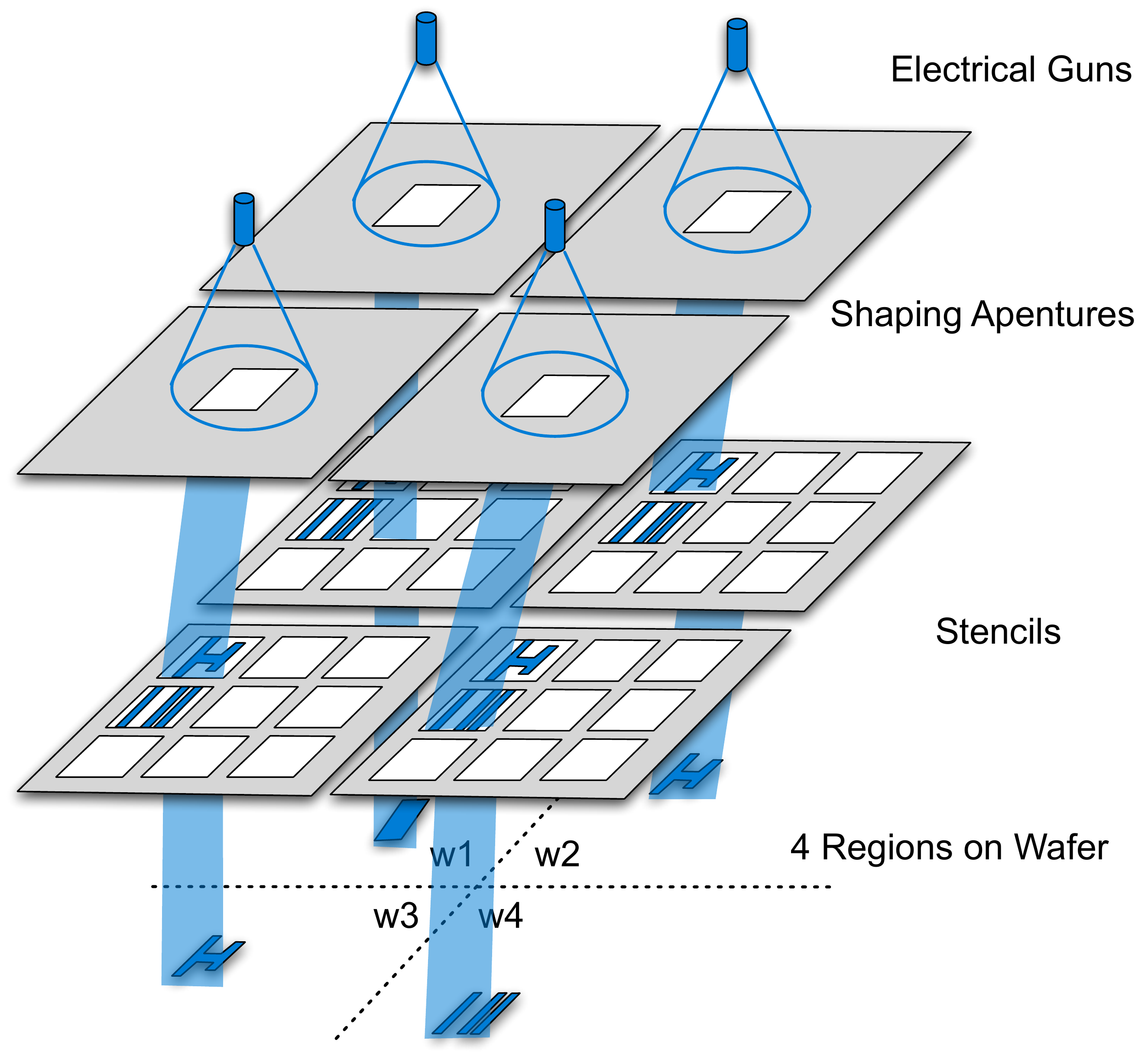}
  \vspace{-.2in}
  \caption{Printing process of MCC system.}
  \label{fig:projectMCC}
  \vspace{-.1in}
\end{figure}

Even with decades of development, the key limitation of the EBL system has been and still is the low throughput.
Recently, multi-column cell (MCC) system is proposed as an extension of conventional EBL system,
where several independent character projections (CP) are used to further speed-up the writing process \cite{EBL_SPIE04_Yasuda,EBL_SPIE2012_Maruyama}.
Each CP is applied on one section of wafer, and all CPs can work parallelly to achieve better throughput.
Due to the design complexity and cost consideration, different CPs share one stencil design \cite{EBL_SPIE2010_Shoji}.
One example of MCC printing process is illustrated in Fig. \ref{fig:projectMCC}, where four CPs are bundled to generate a MCC system.
The whole wafer is divided into four regions, $w_1, w_2, w_3$ and $w_4$, and each region is printed through one CP.
Note that the whole writing time of the MCC system is determined by the maximum writing time of the four regions.
For modern design, because of the numerous distinct circuit patterns,  only limited number of patterns can be employed on stencil.
Therefore, the area constraint of stencil is the bottleneck. 
To improve the throughput, the stencil should be carefully designed/manufactured to contain the most repeated cells or patterns.

Much previous works focus on the design optimization for conventional EBL system
\cite{EBL_IEICE06_Sugihara,EBL_ISPD2011_Yuan,EBL_TCAD2012_Yuan,EBL_ASPDAC2012_Du,EBL_ASPDAC2013_Yu}.
Stencil planning, which is one of the most challenges, has earned much attentions.
When blank overlapping is not considered, the stencil planning can be formulated as a character selection problem,
where integer linear programming (ILP) was applied to select group of characters for throughput maximization \cite{EBL_IEICE06_Sugihara}. 
Recently, Yuan et al. in \cite{EBL_TCAD2012_Yuan}
investigated on the overlapping aware stencil planning (OSP) problem.


However, no existing stencil planning work has done for the MCC system.
Compared with conventional EBL system, MCC system introduces two main challenges.
First, the objective is new: in MCC system the wafer is divided into several regions, and each region is written by one CP. Therefore the new OSP should minimize the maximal writing times of all regions.
While in conventional EBL system, the objective is simply minimize the wafer writing time.
Besides, the stencil for an MCC system can contain more than 4000 characters, previous methodologies for EBL system may suffer from runtime penalty.

\begin{figure} [tb]
  \centering
  \vspace{-.1in}
    \subfigure[]{
      \includegraphics[width=0.18\textwidth]{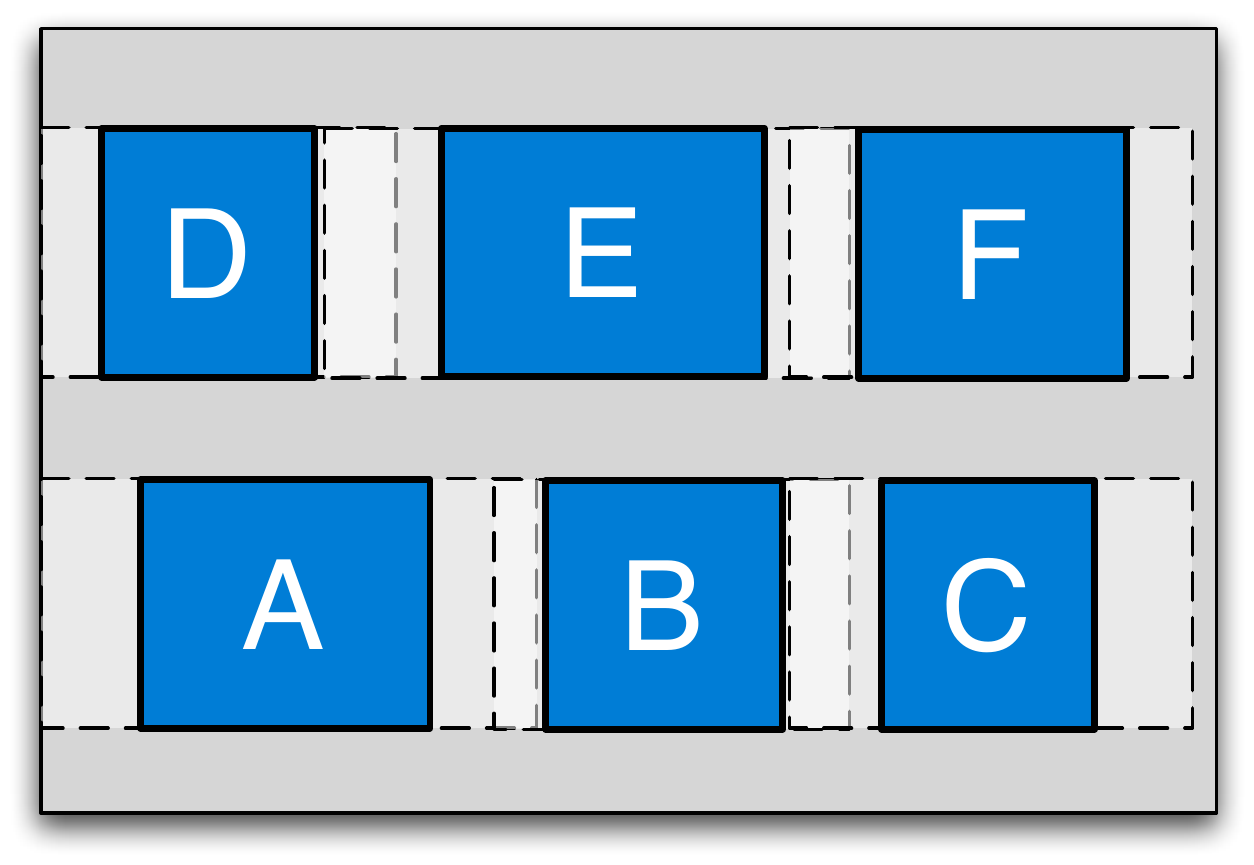}
    }
   \hspace{.2em}
   \subfigure[]{
      \includegraphics[width=0.18\textwidth]{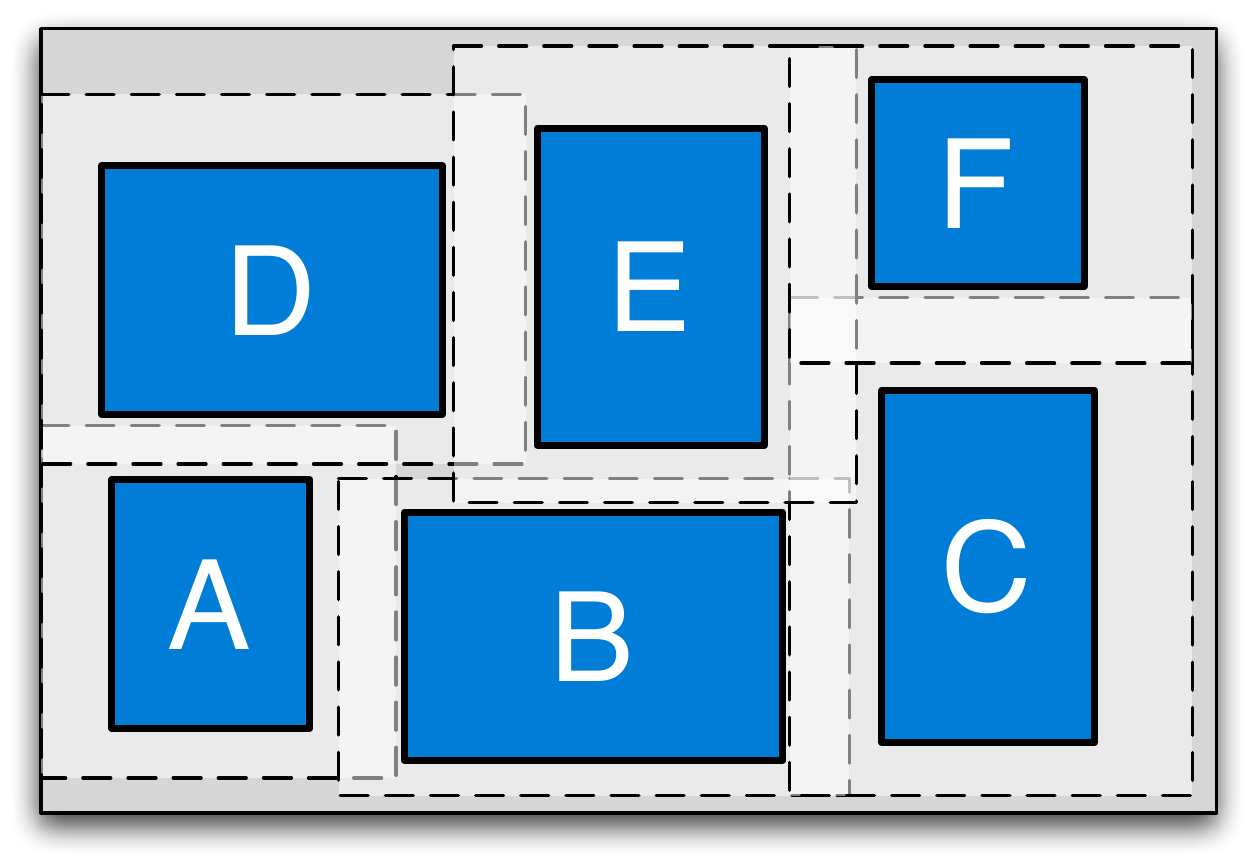}
    }
  \vspace{-.1in}
  \caption{(a) 1D-OSP. (b) 2D-OSP.}
  \label{fig:OSP}
  \vspace{-.1in}
\end{figure}

The OSP problem can be divided into two sub-problems: \textit{1D-OSP} and \textit{2D-OSP} \cite{EBL_TCAD2012_Yuan}.
When standard cells with same height are selected into stencil, the problem is referred as 1D-OSP.
As shown in Fig.~\ref{fig:OSP}(a), each character implements one standard cell, and the enclosed circuit patterns of all the characters have the same height.
Note that here we only show the horizontal blanks, and the vertical blanks are not represented because they are identical.
In 2D-OSP, the blanking spaces of characters are non-uniform along both horizontal and vertical directions.
By this way, stencil can contain both complex via patterns and regular wires.
Fig.~\ref{fig:OSP}(b) illustrates a stencil design example for 2D-OSP.

This paper presents E-BLOW, the first study for OSP problem in MCC system.
E-BLOW integrates several novel techniques to achieve near-optimal solution in reasonable time.
The main contributions of this paper are stated as follows:
(1) First study for stencil planning problem in MCC system.
(2) Shows that OSP problem for both EBL and MCC systems are NP-hard.
(3) Proposes a simplified formulation for 1D-OSP, and proves its rounding lower bound theoretically.
(4) A successive relaxation algorithm to find a near optimal solution.
(5) KD-Tree based clustering algorithm for speedup in 2D-OSP.

The remainder of this paper is organized as follows.
Section \ref{sec:problem} provides problem formulation.
Section \ref{sec:1d} presents algorithm details to resolve 1D-OSP problem in E-BLOW,
while section \ref{sec:2d} details the E-BLOW solutions to 2D-OSP problem.
Section \ref{sec:result} reports experimental results, followed by the conclusion in Section \ref{sec:conclusion}.

%% file: doc/problem.tex
\vspace{-.05in}
\section{Problem Formulation}
\label{sec:problem}

\vspace{-.05in}
\subsection{OSP Problem Formulation}

In an MCC system with $P$ CPs, the whole wafer is divided into $P$ regions $\{w_1, w_2, \dots, w_P\}$,
and each region is written by one particular CP.
We assume cell extraction \cite{EBL_SPIE09_Manakli} has been resolved first.
In other words, a set of character candidates $C^C = \{c_1, \cdots, c_n\}$ has already been given to the MCC system.
For each character candidate $c_i \in C^C$, its writing time through VSB mode is denoted as $n_{i}$,
while its writing time through CP mode is $1$.

The regions of wafer have different layout patterns, and the throughputs would be also different.
Suppose character candidate $c_i$ repeats $t_{ic}$ times on region $w_c$.
Let $a_i$ indicate selection of  character candidate $c_i$ as follows.
\begin{displaymath}
	a_i = 
	\left\{
	\begin{array}{cc}
		1, 			& \textrm{candidate } c_i \textrm{ is selected on stencil}\\
		0,  	    & \textrm{otherwise}
	\end{array}
	\right.
\end{displaymath}
If $c_i$ is prepared on stencil, the total writing time of patterns $c_i$ on region $w_c$ is $t_{ic} \cdot 1$.
Otherwise, $c_i$ should be printed through VSB.
Since region $w_c$ comprises $t_{ic}$ candidate $c_i$, the writing time would be $t_{ic} \cdot n_i$.
Therefore, for region $w_c$ the total writing time $T_c$ is as follows:
\begin{eqnarray}
  T_c & = & \sum_{i=1}^{n} a_i \cdot (t_{ic} \cdot 1) + \sum_{i=1}^{n} (1-a_i) \cdot (t_{ic} \cdot n_i) \notag\\
        & = & \sum_{i=1}^n t_{ic} \cdot n_i - \sum_{i=1}^{n} t_{ic}  \cdot (n_i - 1) \cdot a_i = T_{c}^{VSB} - \sum_{i=1}^{n} R_{ic} \cdot a_i \notag
\end{eqnarray}
where $T_{c}^{VSB} = \sum_{i=1}^n t_{ic} \cdot n_i$, and $R_{ic} = t_{ic}  \cdot (n_i - 1)$.
$T_{c}^{VSB}$ represents the writing time on $w_c$ when only VSB is applied, 
and $R_{ic}$ can be viewed as the writing time reduction of candidate $c_i$ on region $w_c$.
In MCC system, both $T_{c}^{VSB}$ and $R_{ic}$ are constants. 
Therefore, the total writing time of the MCC system is formulated as follows:
\begin{figure}[htb]
\vspace{-.2in}
\begin{eqnarray}
  T_{total} & = & \textrm{max} \{T_c\}  \notag\\
                & = & \textrm{max} \{T_{c}^{VSB} - \sum_{i=1}^{n} R_{ic} \cdot a_i \},  \forall c \in P  \label{eq:obj}
\end{eqnarray}
\vspace{-.3in}
\end{figure}

\begin{problem}{Overlapping aware Stencil Planning (OSP) for MCC system}: 
Given a set of character candidate $C^{C}$, select a subset $C^{CP}$ out of $C^{C}$ as characters, and place them on the stencil.
The objective is to minimize the total writing time $T_{total}$ expressed by (\ref{eq:obj}), while the placement of $C^{CP}$ is bounded by the outline of stencil.
The width and height of stencil is $W$ and $H$, respectively.
\end{problem}

For convenience, we use the term OSP to refer OSP for MCC system in the rest of this paper.


\subsection{NP-Hardness}
\vspace{-.1in}

\begin{mylemma}
\label{lem:1d}
1D-OSP problem is NP-hard.
\vspace{-.1in}
\end{mylemma}

Let us consider a special and simper case of 1D-OSP, where each candidate $c_i$ has zero blank space,
and CP number is 1.
Then the problem can be reduced from a multiple knapsack problem,
which is a well known NP-hard problem \cite{book90Knapsack}.

\begin{mylemma}
\label{lem:2d}
2D-OSP problem is NP-hard.
\vspace{-.1in}
\end{mylemma}

Let us consider a special case of 2D-OSP, where each candidate $c_i$ has zero blank space,
and CP number is 1.
The 2D-OSP problem includes two subproblems:
candidate selection and candidate packing.
After some candidates are selected on the stencil, the candidates packing problem can be reduced from a strip packing problem \cite{Packing_ASFCS96}, which is NP-hard.

Combining Lemma \ref{lem:1d} and Lemma \ref{lem:2d}, we can achieve the conclusion that OSP problem, even for conventional EBL system, is NP-hard.

%% file: doc/1D.tex
\section{E-BLOW for 1D-OSP}
\label{sec:1d}

When each character implements one standard cell, the enclosed circuit patterns of all the characters have the same height.
Corresponding OSP problem is called 1D-OSP, which can be viewed as a combination of character selection and single row ordering problems \cite{EBL_TCAD2012_Yuan}.
Different from two heuristic steps proposed in \cite{EBL_TCAD2012_Yuan}, we show that the two problems can be solved simultaneously through ILP formulation (\ref{eq:1ilp}).
\begin{figure}[h]
\vspace{-.2in}
\begin{align}
    \textrm{min} \ & \ \ T_{total}  \label{eq:1ilp}\\
    \textrm{s.t} \ \ 
    & T_{total} \ge T_{c}^{VSB} - \sum_{i=1}^{n}(\sum_{k=1}^{M} R_{ic} \cdot a_{ik}), \ \forall c \in P \label{1ilp_a}\tag{$2a$}\\
    & x_i + w_i \le W,          \qquad \forall i \in N 				\label{1ilp_b}\tag{$2b$}\\
	& \sum_k^m a_{ik} \le 1, \ \qquad \forall k \in M               \label{1ilp_c}\tag{$2c$}\\
    & x_i + w_{ij} - x_j \le W (2 + p_{ij} - a_{ik} - a_{jk})		\label{1ilp_d}\tag{$2d$}\\
    & x_j + w_{ji} - x_i \le W (3 - p_{ij} - a_{ik} - a_{jk})		\label{1ilp_e}\tag{$2e$}\\
    & a_{ik}, a_{jk}, p_{ij}: 0-1 \ \textrm{variable}				\label{1ilp_f}\tag{$2f$}
\end{align}
\vspace{-.2in}
\end{figure}

In (\ref{eq:1ilp}) $W$ is the width constraint of stencil,
$M$ is the number of rows,
and $w_i$ is width of character $c_i$.
$x_i$ is the x-position of $c_i$.
If and only if $c_i$ is assigned to $k$th row, $a_{ik}=1$.
In other words, $a_{ik}$ determines the y-position of $c_i$.
$w_{ij} = w_i -o_{ij}^h$ and $w_{ji} = w_i -o_{ji}^h$, where $o_{ij}^h$ is the overlapping when candidates $c_i$ and $c_j$ are packed together.
Constraints (\ref{1ilp_d}) (\ref{1ilp_e}) are used to check position relationship between $c_i$ and $c_j$.
For $k$th row, it is easy to see that only when $a_{ik} = a_{jk} = 1$, i.e. both character $i$ and character $j$ are assigned to row $j$, then only one of the two constraints (\ref{1ilp_d}) (\ref{1ilp_e}) will be active.
If either of them are not assigned to the row, neither of the constraints are active.
The number of variables for (\ref{eq:1ilp}) is $O(N^2)$, where $N$ is the number of character candidates.

Since ILP is a well known NP-hard problem, directly solving it may suffer from long runtime penalty. 
One straightforward speedup method is to relax ILP (\ref{eq:1ilp}) into linear programming (LP) as following: replacing constraints (\ref{1ilp_f}) by $0 \le a_{ik}, a_{jk}, p_{ij} \le 1$.
It is obvious that the LP solution provides a lower bound to the ILP solution.
However, we observe that the solution of relaxed LP would be like this:
for each $i$, $\sum_j a_{ij}=1$ and all the $p_{ij}$ are assigned $0.5$.
Although the objective function is minimized and all the constraints are satisfied,
this LP relaxation provides no useful information to guide future rounding, i.e.,
all the character candidates are selected and no ordering relationship is determined.

\begin{figure} [tb]
  \centering
  \vspace{-.1in}
  \includegraphics[width=0.4\textwidth]{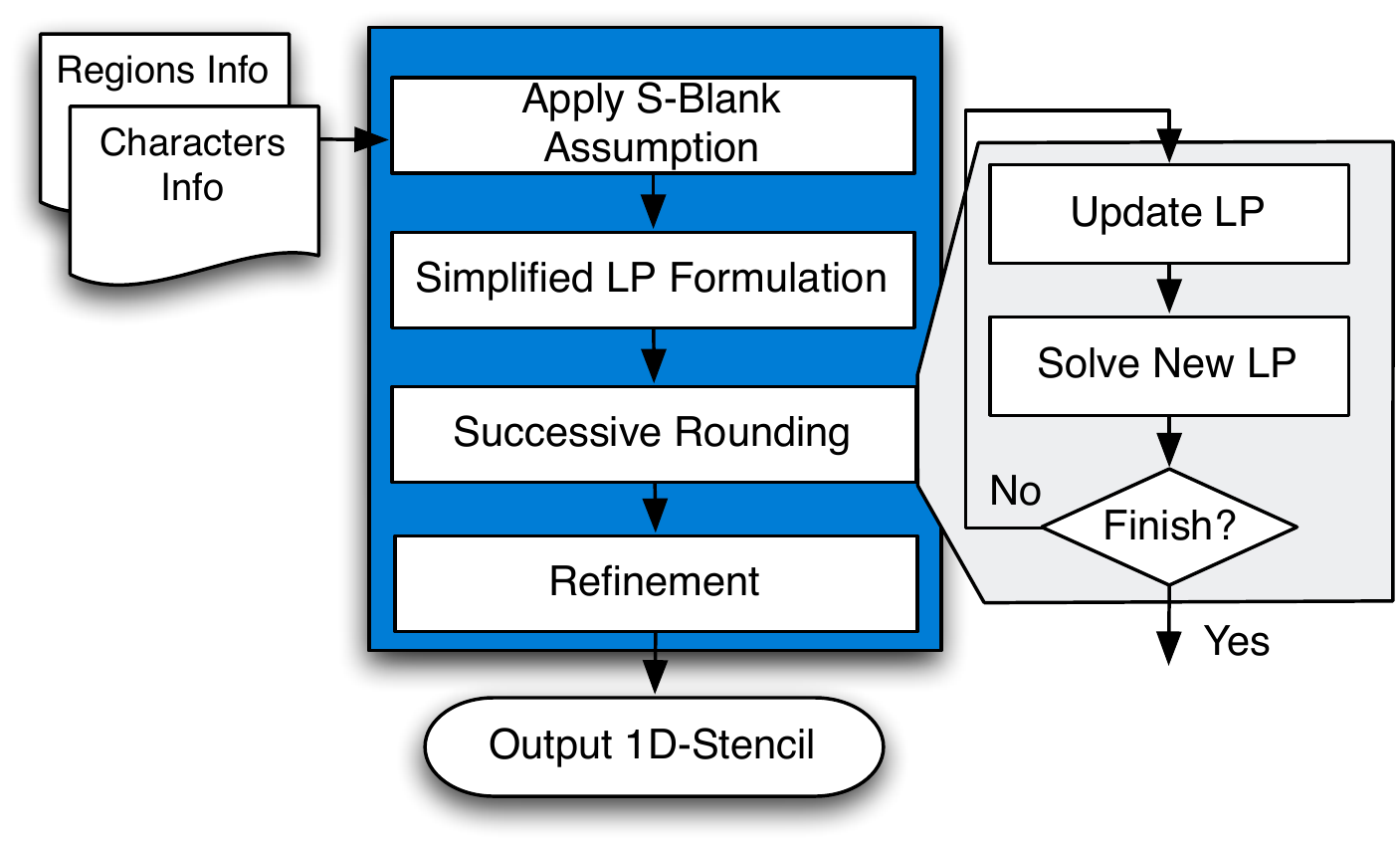}
  \vspace{-.1in}
  \caption{E-BLOW overall flow for 1D-OSP.}
  \label{fig:Flow1D}
  \vspace{-.1in}
\end{figure}

To overcome the limitation of above rounding, E-BLOW proposes a novel iterative solving framework to search near-optimal solution in reasonable runtime.
The main idea is to modify the ILP formulation, so that the corresponding LP relaxation can provide good lower bound theoretically.
As shown in Fig. \ref{fig:Flow1D}, the overall flow includes three parts:
Simplified ILP formulation, Successive Rounding and Refinement.
At section \ref{sec:ilp} the simplified formulation will be discussed, and its LP Rounding lower bound will be proved.
Function \textit{SuccRounding}() is the successive rounding method, which will be introduced at section \ref{sec:succ}.
At last, section \ref{sec:refine} proposes a dynamic programming based refinement.

\subsection{Symmetrical Blank (S-Blank) Assumption}
Our simplified formulation is based on a symmetrical blank assumption:
the blanks of each character is symmetry, left slack equals to right slack.
Note that for different characters $i$ and $j$, their slacks $s_i$ and $s_j$ can be different.

At first glance this assumption may lose optimality, however, it provides several practical and theoretical benefits.
(1) Single row ordering \cite{EBL_TCAD2012_Yuan} was transferred into Hamilton Cycle problem, 
which is a well known NP-hard problem and even particular solver is quite expensive.
Under the assumption, this ordering problem can be optimally solved in $O(n)$.
(2) The ILP formulation can be simplified to provide a reasonable rounding bound theoretically.
Compared with previous heuristic framework \cite{EBL_TCAD2012_Yuan}, the proved rounding bound provides a better guideline for a global view search.
(3) To compensate the inaccuracy in the asymmetrical blank cases, E-BLOW provides a refinement to further improve the throughput.

Given $p$ character candidates, single row ordering problem adjusts the relative locations to minimize the total width. 
Under symmetrical blank assumption, this problem can be optimally solved by a two steps greedy approach.
First, all characters are sorted decreasingly by blanking space $s_i$; second, they are inserted one by one.
Each one can insert at either left end or right end.

\begin{mytheorem}
\label{them:symm}
Under S-Blank assumption, the greedy approach can get maximum overlapping space $\sum_i s_i - \textrm{max}\{s_i\}$.
\vspace{-.1in}
\end{mytheorem}

In practical, we set $s_i = \lceil (sl_i + sr_i ) / 2 \rceil$, where $sl_i$ and $sr_i$ are $c_i$'s left slack and right slack, respectively.

\subsection{Simplified Formulation}
\label{sec:ilp}

To further simplify (\ref{eq:1ilp}), we modify the objective function through assigning each character $c_i$ with one profit value $profit_i$.
Then based on the Theorem \ref{them:symm}, the formulation (\ref{eq:1ilp}) can be simplified as follow:
\begin{align}
    \textrm{max}      & \ \sum_i \sum_j  a_{ij} \cdot profit_i            \label{fast}\\
    \textrm{s.t.} \ \ & \sum_i (w_i - s_i) \cdot a_{ij} \le W - B_j, \forall j       \label{fast_a}\tag{$3a$}\\
                      & B_j \ge s_i \cdot a_{ij} , \forall i                         \label{fast_b}\tag{$3b$}\\
                      & \sum_j a_{ij}  \le 1, \ \ \forall c_i \in C^C                \label{fast_c}\tag{$3c$}\\
                      & a_{ij} = 0 \ \ \textrm{or}\ \ 1                              \label{fast_d}\tag{$3d$}
\end{align}
(\ref{fast_a}) and (\ref{fast_b}) are based on Theorem \ref{them:symm} to calculate the row width, where (\ref{fast_b}) is to linearize $max$ operation.
Here $B_j$ can be viewed as the maximum blank space of all the characters on row $r_j$.
(\ref{fast_c}) means each character can be assigned into at most one row.
It's easy to see that the number of variables is $O(nm)$.
Generally speaking, single character number $n$ is much larger than row number $m$, so compared with basic ILP formulation (\ref{eq:1ilp}), the variable number of (\ref{fast}) can be reduced dramatically.

Furthermore, theoretically the simplified formulation (\ref{fast}) can achieve reasonable LP rounding lower bound.
To explain this, let us first look at a similar program (\ref{knapsack}) as follows:
\begin{align}
    \textrm{max}     & \ \sum_i \sum_j  (w_i - s_i) \cdot a_{ij} \cdot ratio_i                 \label{knapsack}\tag{$3'$}\\
    \textrm{s.t.} \ \ & \sum_i (w_i - s_i) \cdot a_{ij} \le W - max_s      	              \label{knapsack_a}\tag{$3a'$}\\
                                  &  (3c) - (3d)  \notag
\end{align}
where $ratio_i = profit_i / (w_i - s_i)$,
and $max_s$ is the maximum horizontal slack length of every character, i.e. $max_s = \textrm{max}\{s_i | i = 1, 2, \dots, n\}$.
Program (\ref{knapsack}) is a well known multiple knapsack problem \cite{book90Knapsack}.


\begin{mylemma}
\label{lem:2}
If each $ratio_i$ is the same, the multiple knapsack problem (\ref{knapsack}) can find a $1/2-$approximation algorithm using LP Rounding method.
\vspace{-.1in}
\end{mylemma}

For brevity we omit the proof, detailed explanations can be found in \cite{Knapsack_JCO00Dawande}.
It shall be noted that if all $ratio_i$ are the same, program (\ref{knapsack}) can be approximated to a max-flow problem.
Based on Lemma \ref{lem:2}, if we denote $\alpha$ as $\textrm{max} \{ratio_i\}$ /$\textrm{min} \{ratio_i\}$, we can achieve the following theorem:

\begin{mytheorem}
\label{them:3}
The LP Rounding solution of (\ref{fast}) can be a $0.5/\alpha-$ approximation to program (\ref{knapsack}).
\vspace{-.1in}
\end{mytheorem}

Due to space limit, the detailed proof is omitted.
The only difference between (\ref{fast}) and (\ref{knapsack}) is that the right side values at (\ref{fast_a}) and (\ref{knapsack_a}).
Blank spacing is relatively small comparing with the row length, we can get that $W - max_s \approx W - B_j$.
Then based on Theorem \ref{them:3}, we can conclude that program $(3)$ has a reasonable rounding bound.

\vspace{-.05in}
\subsection{Successive Relaxation}
\label{sec:succ}

Because of the reasonable LP rounding property shown in Theorem \ref{them:3}, we propose a successive relaxation algorithm to solve program (\ref{fast}) iteratively.
The ILP formulation (\ref{fast}) becomes an LP if we relax the discrete constraint to a continuous constraint as:
$0 \le a_{ij} \le 1$.
The successive relaxation algorithm is shown in Algorithm \ref{alg:round}.
At first we set all $a_{ij}$ to variables since any $a_{ij}$ is not guided to rows.
The LP is updated and solved iteratively.
For each new LP solution, we search the maximal $a_{pq}$ (line 6).
Then for all $a_{ij}$ that is close the the maximal value $a_{pq}$, we try to pack $c_i$ into row $r_j$, and set it as non-variable.
Note that since several $a_{ij}$ are assigned permanent value, the number of variables in updated LP formulation would continue to decrease.
This procedure repeats until no appropriate $a_{ij}$ can be found.
One key step of the Algorithm \ref{alg:round} is the $profit_i$ update (line 3).
For each character $c_i$, we set its $profit_i$ as follows:
\begin{equation}
  profit_i = \sum_c \frac{t_{c}}{t_{max}} \cdot (n_i - 1) \cdot t_{ic} \label{eq:profit}
\end{equation}
where $t_c$ is current writing time of region $w_c$, and $t_{max} =$ max $\{t_c, \forall c \in P\}$.
Through applying the $profit_i$, the region $w_c$ with longer writing time would be considered more during the LP formulation. 
During successive relaxation, if $c_i$ hasn't been assigned to any row, $profit_i$ would continue to updated, so that the total writing time of the whole MCC system can be minimized.

\begin{algorithm}[tb]
\caption{SuccRounding( $th_{inv}$ )}
\label{alg:round}
\begin{algorithmic}[1]
    \REQUIRE{ ILP Formulation (\ref{fast})}
    \STATE  {set all $a_{ij}$ to variables;}
    \REPEAT
        \STATE {update $profit_i$ for all variables $a_{ij}$;}
        \STATE  {solve relaxed LP of (\ref{fast});}
        \REPEAT
            \STATE {find $a_{pq} = $ max$\{a_{ij}$, and $c_i$ can insert into row $r_j$\};}
            \FORALL {$a_{ij} \ge a_{pq} \times th_{inv}$ }
              \IF {$c_i$ can be assigned to row $r_j$}
                \STATE { $a_{ij} = 1$ and set it to a non-variable; }
                \STATE {Update capacity of row $r_j$;}
              \ENDIF
            \ENDFOR
         \UNTIL  {cannot find $a_{pq}$}
    \UNTIL  { }
\end{algorithmic}
\end{algorithm}

\vspace{-.05in}
\subsection{Refinement}
\label{sec:refine}

Simplified formulation and successive relaxation are under the symmetrical blank assumption.
Although it can be effectively solved, for asymmetrical cases it would lose some optimality.
To compensate the losing, we present a dynamic programming based refinement procedure.
As discussed above, for $k$ characters, single row ordering can have $2^{k-1}$ possible solutions.
Under symmetrical blank space assumption, all these orderings get the same length.
But for the asymmetrical cases, it does not hold anymore.
Our dynamic programming based algorithm \textit{Refine}(k) finds the best solution from these $2^{k-1}$ options.
The detailed is shown in Algorithm \ref{alg:refine}.
At first, if $k > 1$, then \textit{Refine}(k) will recursively call \textit{Refine}(k-1) to generate all old partial solutions.
All these partial solutions will be updated by adding candidate $c_k$ (lines 6-8).
Note that maintaining all solutions is impractical and unnecessary, because many of them are inferior to others.
In \textit{SolutionPruning}(), all solutions are checked, if one solution $S_A$ is inferior to another solution $S_B$, $S_A$ would be pruned to save computation cost.
For each solution a triplet $(w, l, r)$ is constructed to store the information of width, left slack and right slack.
We define the \textit{inferior} relationship as follow.
For two solutions  $S_A = (w_a, l_a, r_a)$ and $S_B = (w_b, l_b, r_b)$, $S_B$ is inferior to $S_A$ if and only if $w_a \ge w_b$, $l_a \le l_b$ and $r_a \le r_b$.

\begin{algorithm} [htb]
\caption{Refine(k)}
\label{alg:refine}
\begin{algorithmic}[1]
  \IF { k = 1 }
      \STATE Generate partial solution $(w_1, sl_1, sr_1)$;
  \ELSE
      \STATE Refine(k-1);
      \FOR{ each partial solution $(w, l, r)$}
          \STATE $(w_1, l_1, r_1) = (w+w_k-\textrm{min}(sr_k, l), sl_k, r)$;
          \STATE $(w_2, l_2, r_2) = (w+w_k-\textrm{min}(sl_k, r), l, sr_k)$;
          \STATE Replace $(w, l, r)$ by $(w_1, l_1, r_1)$ and $(w_2, l_2, r_2)$;
          \IF {solution set size $\ge$ threshold}
              \STATE SolutionPruning();
          \ENDIF
      \ENDFOR
  \ENDIF
\end{algorithmic}
\end{algorithm}

After \textit{Refine(k)} for each row, if more available spaces are generated, a greedy insertion approach similar to \cite{EBL_TCAD2012_Yuan} would be proposed to further improve the throughput.

%% file: doc/2D.tex
\section{E-BLOW  for 2D-OSP}
\label{sec:2d}

\begin{figure} [hb]
  \centering
  \vspace{-.1in}
  \includegraphics[width=0.3\textwidth]{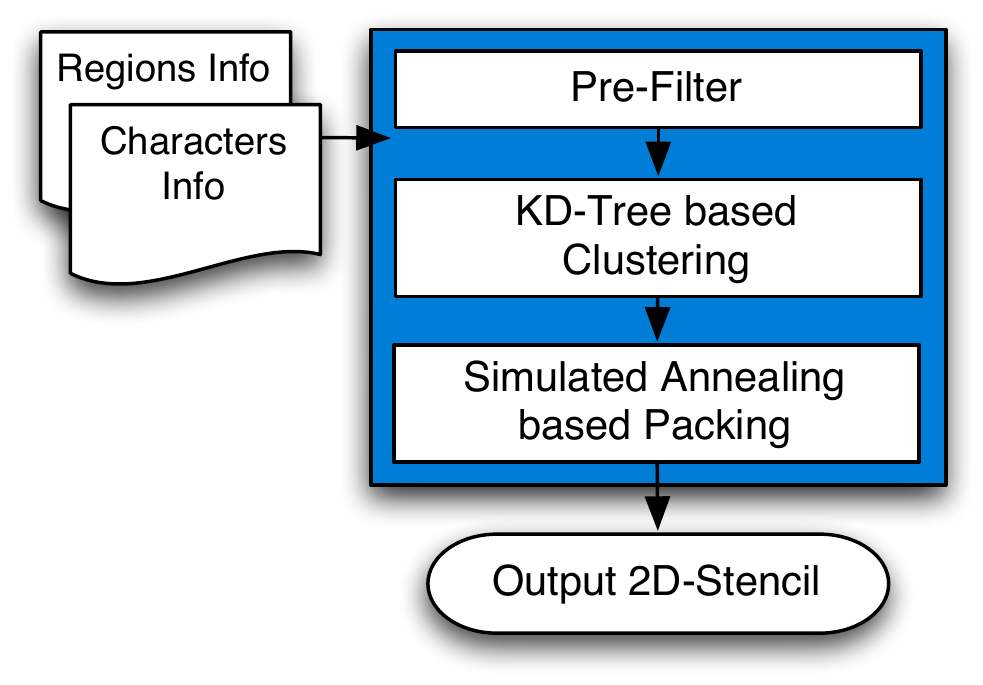}
  \vspace{-.1in}
  \caption{E-BLOW overall flow for 2D-OSP.}
  \label{fig:Flow2D}
  \vspace{-.1in}
\end{figure}

Now we consider a more general case: the blanking spaces of characters are non-uniform along both horizontal and vertical directions. 
This problem is referred as 2D-OSP problem.
In \cite{EBL_TCAD2012_Yuan} the 2D-OSP problem was transformed into a floorplanning problem.
However, several key differences between traditional floorplanning and OSP were ignored.
(1) In OSP there is no wirelength to be considered, while at floorplanning wirelength is a major optimization objective.
(2) Compared with complex IP cores, lots of characters may have similar sizes.
(3) Traditional floorplanner could not handle the problem size of modern MCC design. 
To deal with all these properties, an approximation packing framework is proposed (see Fig. \ref{fig:Flow2D}).
Given the input character candidates, the pre-filter process is first applied to remove characters with bad profit (defined in (\ref{eq:profit})).
Then the second step is a KD-Tree based clustering algorithm to effectively speed-up the design process.
Followed by the final floorplanner to pack all candidates.

\subsection{KD-Tree based Clustering}

\begin{algorithm}[htb]
\caption{KD-Tree based Clustering}
\label{alg:cluster}
\begin{algorithmic}[1]
  \REQUIRE{ set of candidates $C^C$. }
     \REPEAT
	\STATE  { Sort all candidates by $profit_i$;}
	\STATE  { Set each candidates $c_i$ to unclustered;}
	\FORALL {unclustered candidate $c_i$}
	    \STATE Find pair $(c_i, c_j)$ with similar blank spaces and profits;
	    \STATE Cluster $(c_i, c_j)$, label them as clustered;
	\ENDFOR
	\STATE  { Update candidate information;}
   \UNTIL  {reach clustering threshold}
\end{algorithmic}
\end{algorithm}

Clustering is a well studied problem, and there are many of works and applications in VLSI
\cite{PAR_VLSIJ95_Alpert}.
However, previous methodologies cannot be directly applied here.
(1) Traditional clustering is based on netlist, which provides the all clustering options.
Generally speaking, netlist is sparse, but in OSP the connection relationships are so complex that any two characters can be clustered, and totally there are $O(n^2)$ clustering options.
(2) Given two candidates $c_i$ and $c_j$, there are several clustering options.
For example, horizontal clustering and vertical clustering may have different overlapping space.

The details of our clustering procedure are shown in Algorithm \ref{alg:cluster}.
The clustering is repeated until the clustered candidate number reaches the clustering threshold.
Initially all the candidates are sorted by $profit_i$, it means those candidates with more shot number reduction are tend to be clustered. Then clustering (lines 3-8) is carried out.
For each candidate $c_i$, finding available $c_j$ may need $O(n)$, and complexity of the horizontal clustering and vertical clustering are both $O(n^2)$. Then the complexity of the whole procedure is $O(n^2)$, where $n$ is the number of candidates.

\begin{figure} [tb]
  \centering
  \subfigure[]{\includegraphics[width=0.2\textwidth]{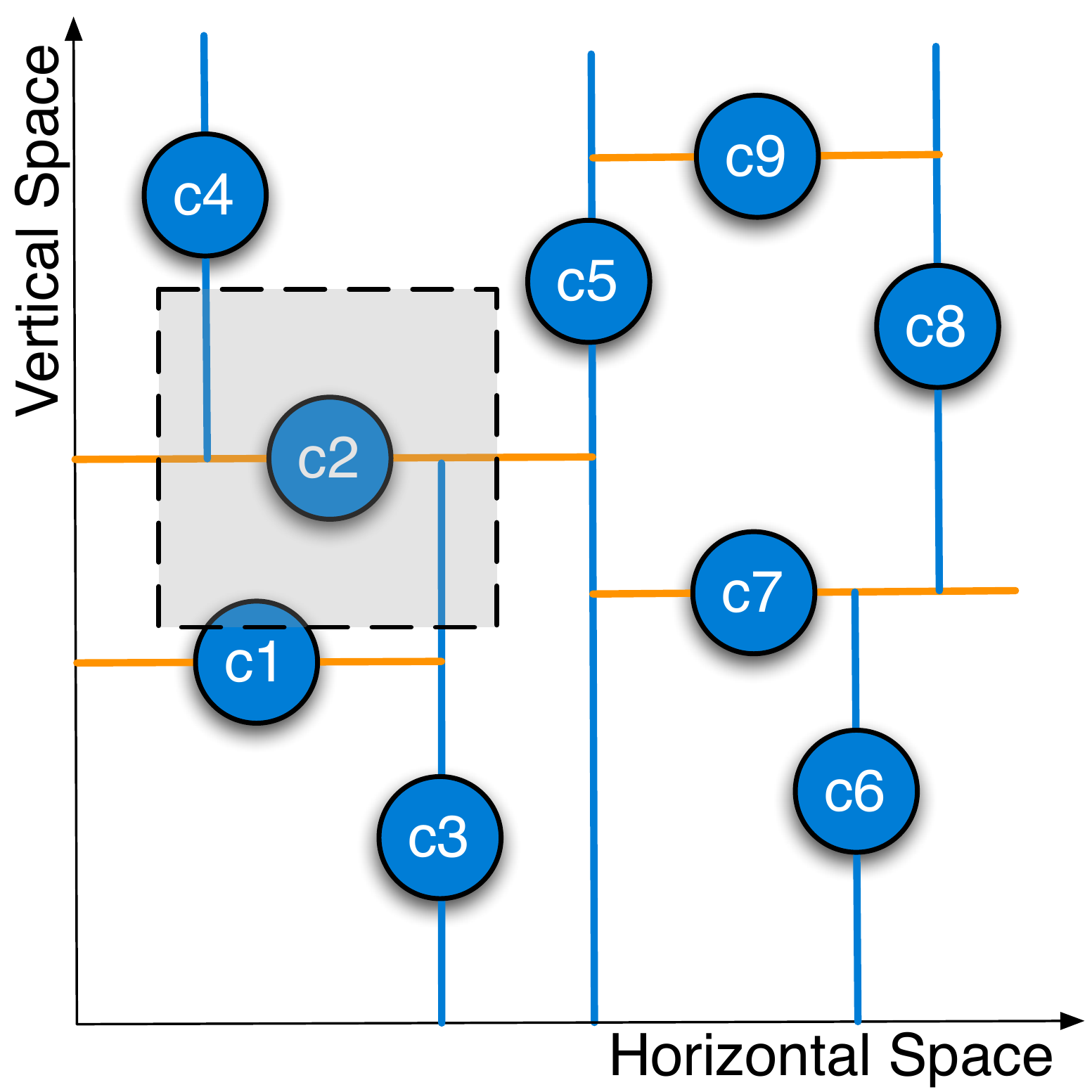}}
  \hspace{-0.1in}
  \subfigure[]{\includegraphics[width=0.22\textwidth]{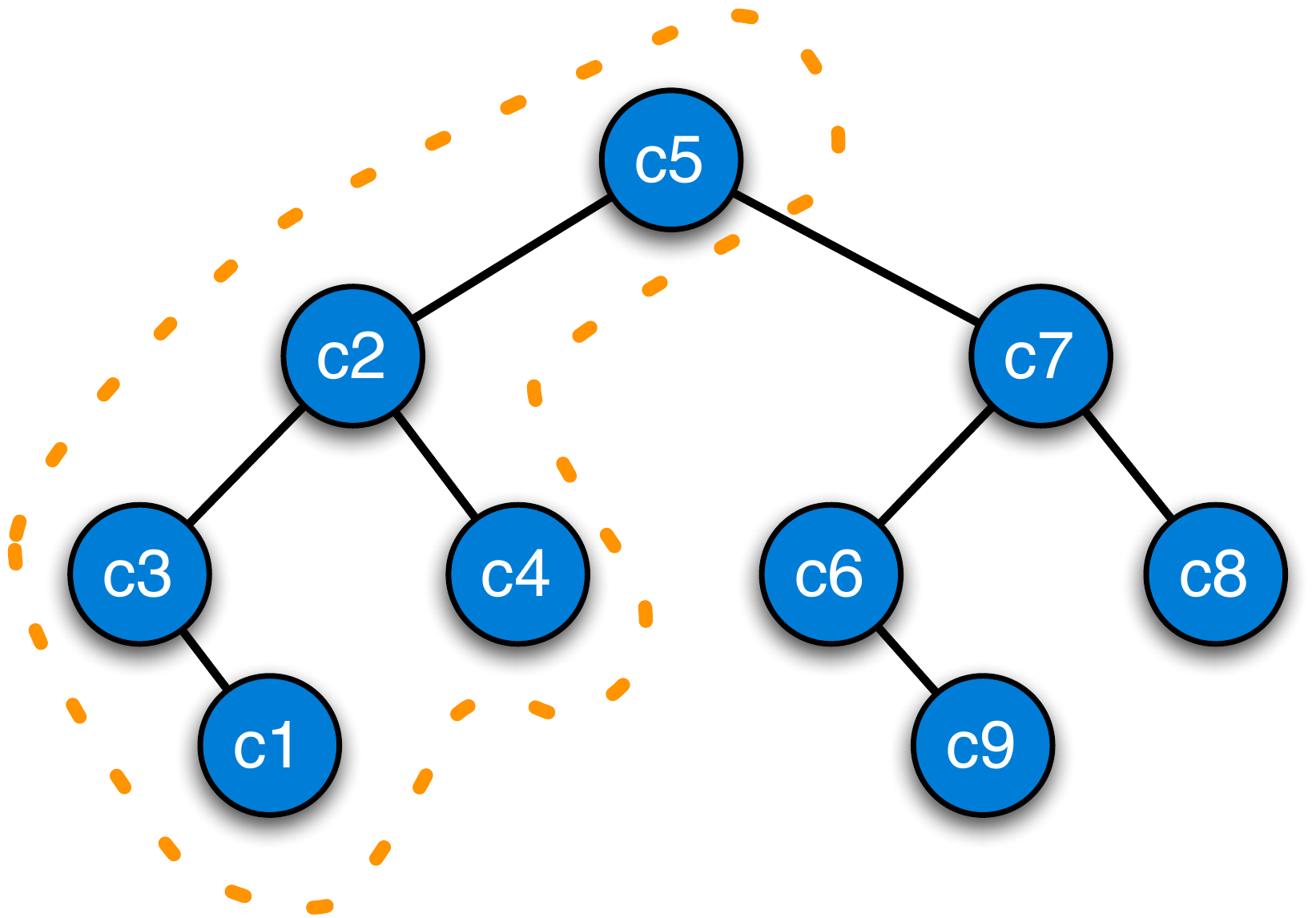}}
  \vspace{-.1in}
  \caption{KD-Tree based region searching.}
  \label{fig:kdtree}
  \vspace{-.1in}
\end{figure}

A KD-Tree~\cite{KDTree} is used to speed-up the process of finding available pair $(c_i, c_j)$.
It provides fast $O(log n)$ region searching operations which keeping the time for insertion and deletion small:
insertion, $O(log n)$; deletion of the root, $O(n(k-1)/k)$; deletion of a random node, $O(log n)$.
Using KD-Tree, the complexity of the Algorithm \ref{alg:cluster} can be reduced to $O(nlog n)$.
A simple example is shown in Fig. \ref{fig:kdtree}.
For the sake of convenience, here for each candidate we only consider horizontal and vertical space.
Given candidate $c_2$, to find another candidate with the similar space, it may need $O(n)$ to scan all other candidates.
However, using the KD-Tree structure shown in Fig. \ref{fig:kdtree}(a), this finding procedure can be viewed as a region searching, which can be resolved in $O(log n)$.
Particularly, as shown in Fig. \ref{fig:kdtree}(b), only candidates $c_1 - c_5$ are scanned.

\subsection{Approximation Framework for 2D-OSP}
\label{subsec:sa}

In E-BLOW we adopt a simulated annealing based framework similar to that in \cite{EBL_TCAD2012_Yuan}.
To demonstrate the effectivity of our pre-filter and clustering methodologies, E-BLOW uses the same parameters with that in \cite{EBL_TCAD2012_Yuan}.
Sequence Pair \cite{FLOOR_TCAD96_SP} is used as a topology representation.

%% file: doc/result.tex
\vspace{-.05in}
\section{Experimental Results}
\label{sec:result}

E-BLOW is implemented in the C++ programming language and executed on a Linux machine with two 3.0GHz CPU and 32GB Memory.
GUROBI~\cite{Gurobi} is used to solve linear programming. 
Eight benchmarks from \cite{EBL_TCAD2012_Yuan} are tested.
Besides, eight benchmarks (1M-x) are designed for 1D-OSP and the other eight (2M-x) are generated for the 2D-OSP problem.
Character projection (CP) number are all set as 10.
For small cases (1M-1, $\dots$, 1M-4, 2M-1, $\dots$, 2M-4) the character candidate number is 1000,
and the stencil size is set as $1000\mu m \times 1000 \mu m$.
For larger cases (1M-5 , $\dots$, 1M-8, 2M-5, $\dots$, 2M-8) the character candidate number is 4000,
and the stencil size is set as $2000\mu m \times 2000 \mu m$.
The size and blank width of each character is similar to that in \cite{EBL_TCAD2012_Yuan}.


\begin{table*}[tb]
\centering 
\caption{Result Comparison for 1D-OSP}
\label{table:1d}
\begin{tabular}{|c|c|c|c|c|c|c|c|c|c|c|c|}
\hline 
\hline 
 &char &CP & \multicolumn{3}{c|}{Greedy in \cite{EBL_TCAD2012_Yuan}} & \multicolumn{3}{c|}{\cite{EBL_TCAD2012_Yuan}}&\multicolumn{3}{c|}{E-BLOW}\\
 \cline{4-12} & \# & \#  & shot \# & char \# & CPU(s) & shot \# & char \# & CPU(s) & shot \# & char \# & CPU(s)\\
 \hline
 1D-1 &1000 &1 	  &79193  &876     &0.2         &50809   &926     &13.5        &29536   &934     &2.2  \\
 1D-2 &1000 &1 	  &122259 &806     &0.2         &93465   &854     &11.8        &44544   &863     &2    \\
 1D-3 &1000 &1 	  &179822 &708     &0.2         &152376  &749     &9.13        &78704   &758     &2.7  \\
 1D-4 &1000 &1 	  &223420 &645     &0.2         &193494  &687     &7.7         &107460  &699     &3.4  \\
                  
 1M-1 &1000 &10   &83786  &876     &0.2         &53333   &926     &13.5        &45243   &938     &4.3  \\
 1M-2 &1000 &10   &123048 &806     &0.2         &95963   &854     &11.8        &81636   &868     &5.4  \\
 1M-3 &1000 &10   &184950 &708     &0.2         &156700  &749     &9.2         &140079  &769     &10.8 \\
 1M-4 &1000 &10   &225468 &645     &0.2         &196686  &687     &7.7         &179890  &707     &7.6  \\
                  
 1M-5 &4000 &10   &377864 &3417    &1.02        &255208  &3629    &1477.3      &227456  &3650    &59.2 \\
 1M-6 &4000 &10   &542627 &315     &1.02        &417456  &3346    &1182        &373324  &3388    &65.1 \\
 1M-7 &4000 &10   &760650 &2809    &1.02        &644288  &2986    &876         &570730  &3044    &58.68\\
 1M-8 &4000 &10   &930368 &2565    &1.01        &809721  &2734    &730.7       &734411  &2799    &65.3 \\
 \hline
 Avg. & - & - 	  &319454.6 &1264.7 &0.47      &259958.3  &1594.0 &362.5    &217751.1 &1618.1 &23.9  \\
 Ratio &-&-	      &\textbf{1.47}     &0.78   &\textbf{0.02}      &\textbf{1.19}      &0.99   &\textbf{15.2}     &\textbf{1.0}        &1.0      &\textbf{1.0}     \\
\hline \hline
\end{tabular}
\end{table*}

\begin{table*}[tb]
\centering 
\caption{Result Comparison for 2D-OSP}
\label{table:2d}
\begin{tabular}{|c|c|c|c|c|c|c|c|c|c|c|c|}
\hline 
\hline 
 &char &CP & \multicolumn{3}{c|}{Greedy in \cite{EBL_TCAD2012_Yuan}} & \multicolumn{3}{c|}{\cite{EBL_TCAD2012_Yuan}}&\multicolumn{3}{c|}{E-BLOW}\\
 \cline{4-12} & \# & \#  & shot \# & char \# & CPU(s) & shot \# & char \# & CPU(s) & shot \# & char \# & CPU(s)\\
 \hline
 2D-1 &1000 &1    &159654  &734  &2.1  	  &107876  &826  &329.6 	&105723  &789  &65.5 \\
 2D-2 &1000 &1    &269940  &576  &2.4 	  &166524  &741  &278.1 	&170934  &657  &52.5 \\
 2D-3 &1000 &1    &290068  &551  &2.6  	  &210496  &686  &296.7 	&178777  &663  &56.4 \\
 2D-4 &1000 &1    &327890  &499  &2.7  	  &240971  &632  &301.7 	&179981  &605  &54.7 \\
 2M-1 &1000 &10   &168279  &734  &2.1  	  &122017  &811  &313.7 	&91193   &777  &58.6 \\
 2M-2 &1000 &10   &283702  &576  &2.4 	  &187235  &728  &286.1 	&163327  &661  &48.7 \\
 2M-3 &1000 &10   &298813  &551  &2.6  	  &235788  &653  &289   	&162648  &659  &52.3 \\
 2M-4 &1000 &10   &338610  &499  &2.7  	  &270384  &605  &285.6 	&195469  &590  &53.3 \\
 2M-5 &4000 &10   &824060  &2704 &19   	  &700414  &2913 &3891  	&687287  &2853 &59   \\
 2M-6 &4000 &10   &1044161 &2388 &20.2	  &898530  &2624 &4245  	&717236  &2721 &60.7 \\
 2M-7 &4000 &10   &1264748 &2101 &21.9 	  &1064789 &2410 &3925.5	&921867  &2409 &57.1 \\
 2M-8 &4000 &10   &1331457 &2011 &22.8 	  &1176700 &2259 &4550.0	&1104724 &2119 &57.7 \\
 \hline
 Avg. & -   & -   &550115  &1218.1&8.3    &448477  &1324 &1582.7    &389930.5&1291.9&56.375\\
 Ratio&-    & -   &\textbf{1.41}    &0.94  &\textbf{0.15}   &\textbf{1.15}    &1.02 &\textbf{28.1}      &\textbf{1.0}       &1.0     &\textbf{1.0}   \\
\hline \hline
\end{tabular}
\end{table*}

\vspace{-.05in}
\subsection{Comparison for 1D-OSP}

For 1D-OSP, Table \ref{table:1d} compares E-BLOW with greedy method and the heuristic framework in \cite{EBL_TCAD2012_Yuan}.
Note that the greedy method was also described in \cite{EBL_TCAD2012_Yuan}.
Column "char \#" is number of character candidates, and column ``CP\#" is number of character projections.
For each algorithm, we record ``shot \#", ``char \#" and ``CPU(s)",
where ``shot \#" is final number of shots and ``char \#" is number of characters on final stencil, ``CPU(s)" reports the runtime.
From table \ref{table:1d} we can see E-BLOW achieve better performance and runtime.
Compared with E-BLOW, the greedy algorithm introduces $47\%$ more shots number, and \cite{EBL_TCAD2012_Yuan} would introduce $19\%$ more shots number.
Note that compared with heuristic method in \cite{EBL_TCAD2012_Yuan}, mathematical formulation can provide global view, even for traditional EBL system (1D-1, $\dots$, 1D-4), E-BLOW achieves better shot number.
Besides, E-BLOW can reduce 34.3\% of runtime.

\vspace{-.05in}
\subsection{Comparison for 2D-OSP}
For 2D-OSP, Table \ref{table:2d} gives the similar comparison.
For each algorithm, we also record ``shot \#", ``char \#" and ``CPU(s)",
where the meanings are the same with that in Table \ref{table:1d}.
From the table we can see that for each test case, although the greedy algorithm is faster, its design results are not good that it would introduce 30\% more shot number.
Besides, compared with the work in \cite{EBL_TCAD2012_Yuan}, E-BLOW can achieve better performance that the shot number can be reduced by 14\%.
Meanwhile, because of  the clustering method, E-BLOW can reach $2.8\times$ speed-up.

From both tables we can see that compared with \cite{EBL_TCAD2012_Yuan}, E-BLOW can achieve a better tradeoff between runtime and performance. 

%% file: doc/conclu.tex
\vspace{-.1in}
\section{Conclusion}
\label{sec:conclusion}

In this paper, we have proposed E-BLOW, a tool to solve OSP problem in MCC system.
For 1D-OSP, a successive relaxation algorithm and a dynamic programming based refinement are proposed.
For 2D-OSP, a KD-Tree based clustering method is integrated into simulated annealing framework.
Experimental results show that compared with previous works, E-BLOW can achieve better performance in terms of shot number and runtime, for both MCC system and traditional EBL system.
As EBL, including MCC system, are widely used for mask making and also gaining momentum for direct wafer writing, we believe a lot more research can be done for not only stencil planning, but also EBL aware design.

\section*{Acknowledgment}

This work is supported in part by NSF and NSFC.

%% file: DAC13_EBLOW.bbl
\begin{thebibliography}{10}
\providecommand{\url}[1]{#1}
\csname url@samestyle\endcsname
\providecommand{\newblock}{\relax}
\providecommand{\bibinfo}[2]{#2}
\providecommand{\BIBentrySTDinterwordspacing}{\spaceskip=0pt\relax}
\providecommand{\BIBentryALTinterwordstretchfactor}{4}
\providecommand{\BIBentryALTinterwordspacing}{\spaceskip=\fontdimen2\font plus
\BIBentryALTinterwordstretchfactor\fontdimen3\font minus
  \fontdimen4\font\relax}
\providecommand{\BIBforeignlanguage}[2]{{%
\expandafter\ifx\csname l@#1\endcsname\relax
\typeout{** WARNING: IEEEtran.bst: No hyphenation pattern has been}%
\typeout{** loaded for the language `#1'. Using the pattern for}%
\typeout{** the default language instead.}%
\else
\language=\csname l@#1\endcsname
\fi
#2}}
\providecommand{\BIBdecl}{\relax}
\BIBdecl

\bibitem{LITH_ICCAD2012_Yu}
B.~Yu, J.-R. Gao, D.~Ding, Y.~Ban, J.-S. Yang, K.~Yuan, M.~Cho, and D.~Z. Pan,
  ``Dealing with {IC} manufacturability in extreme scaling,'' in \emph{IEEE/ACM
  International Conference on Computer-Aided Design (ICCAD)}, 2012, pp.
  240--242.

\bibitem{DPL_ICCAD08_Kahng}
A.~B. Kahng, C.-H. Park, X.~Xu, and H.~Yao, ``Layout decomposition for double
  patterning lithography,'' in \emph{IEEE/ACM International Conference on
  Computer-Aided Design (ICCAD)}, 2008, pp. 465--472.

\bibitem{TPL_ICCAD2011_Yu}
B.~Yu, K.~Yuan, B.~Zhang, D.~Ding, and D.~Z. Pan, ``Layout decomposition for
  triple patterning lithography,'' in \emph{IEEE/ACM International Conference
  on Computer-Aided Design (ICCAD)}, 2011, pp. 1--8.

\bibitem{TPL_SPIE2012_Lucas}
K.~Lucas, C.~Cork, B.~Yu, G.~Luk-Pat, B.~Painter, and D.~Z. Pan, ``Implications
  of triple patterning for 14 nm node design and patterning,'' in \emph{Proc.
  of SPIE}, vol. 8327, 2012.

\bibitem{EUV_SPIE2010_Arisawa}
Y.~Arisawa, H.~Aoyama, T.~Uno, and T.~Tanaka, ``{EUV} flare correction for the
  half-pitch 22nm node,'' in \emph{Proc. of SPIE}, vol. 7636, 2010.

\bibitem{EBL_SPIE09_Pfeiffer}
H.~C. Pfeiffer, ``New prospects for electron beams as tools for semiconductor
  lithography,'' in \emph{Proc. of SPIE}, 2009.

\bibitem{EBL_SPIE04_Yasuda}
H.~Yasuda, T.~Haraguchi, and A.~Yamada, ``A proposal for an {MCC} (multi-column
  cell with lotus root lens) system to be used as a mask-making e-beam tool,''
  in \emph{Proc. of SPIE}, 2004.

\bibitem{EBL_SPIE2012_Maruyama}
T.~Maruyama, Y.~Machida, S.~Sugatani, H.~Takita, H.~Hoshino, T.~Hino, M.~Ito,
  A.~Yamada, T.~Iizuka, S.~Komatsue, M.~Ikeda, and K.~Asada, ``{CP} element
  based design for 14nm node {EBDW} high volume manufacturing,'' in \emph{Proc.
  of SPIE}, 2012.

\bibitem{EBL_SPIE2010_Shoji}
M.~Shoji, T.~Inoue, and M.~Yamabe, ``Extraction and utilization of the
  repeating patterns for {CP} writing in mask making,'' in \emph{Proc. of
  SPIE}, 2010.

\bibitem{EBL_IEICE06_Sugihara}
M.~Sugihara, T.~Takata, K.~Nakamura, R.~Inanami, H.~Hayashi, K.~Kishimoto,
  T.~Hasebe, Y.~Kawano, Y.~Matsunaga, K.~Murakami, and K.~Okumura, ``Cell
  library development methodology for throughput enhancement of character
  projection equipment,'' \emph{IEICE Transactions on Electronics}, vol. E89-C,
  pp. 377--383, 2006.

\bibitem{EBL_ISPD2011_Yuan}
K.~Yuan and D.~Z. Pan, ``{E-B}eam lithography throughput improvement with
  stencil planning and optimization,'' in \emph{ACM International Symposium on
  Physical Design (ISPD)}, 2011.

\bibitem{EBL_TCAD2012_Yuan}
K.~Yuan, B.~Yu, and D.~Z. Pan, ``{E-B}eam lithography stencil planning and
  optimization with overlapped characters,'' \emph{IEEE Transactions on
  Computer-Aided Design of Integrated Circuits and Systems (TCAD)}, vol.~31,
  no.~2, pp. 167--179, Feb. 2012.

\bibitem{EBL_ASPDAC2012_Du}
P.~Du, W.~Zhao, S.-H. Weng, C.-K. Cheng, and R.~Graham, ``Character design and
  stamp algorithms for character projection electron-beam lithography,'' in
  \emph{IEEE/ACM Asia and South Pacific Design Automation Conference (ASPDAC)},
  2012.

\bibitem{EBL_ASPDAC2013_Yu}
B.~Yu, J.-R. Gao, and D.~Z. Pan, ``{L-Shape} based layout fracturing for e-beam
  lithography,'' in \emph{IEEE/ACM Asia and South Pacific Design Automation
  Conference (ASPDAC)}, 2013.

\bibitem{EBL_SPIE09_Manakli}
S.~Manakli, H.~Komami, M.~Takizawa, T.Mitsuhashi, and L.~Pain, ``Cell
  projection use in mask-less lithography for 45nm \& 32nm logic nodes,'' in
  \emph{Proc. of SPIE}, 2009.

\bibitem{book90Knapsack}
S.~Martello and P.~Toth, \emph{Knapsack problems: algorithms and computer
  implementations}.\hskip 1em plus 0.5em minus 0.4em\relax New York, NY, USA:
  John Wiley \& Sons, Inc., 1990.

\bibitem{Packing_ASFCS96}
C.~Kenyon and E.~Remila, ``Approximate strip packing,'' in \emph{Foundations of
  Computer Science, 1996. Proceedings., 37th Annual Symposium on}, oct 1996,
  pp. 31 --36.

\bibitem{Knapsack_JCO00Dawande}
M.~Dawande, J.~Kalagnanam, P.~Keskinocak, F.~Salman, and R.~Ravi,
  ``Approximation algorithms for the multiple knapsack problem with assignment
  restrictions,'' \emph{Journal of Combinatorial Optimization}, vol.~4, pp.
  171--186, 2000.

\bibitem{PAR_VLSIJ95_Alpert}
C.~J. Alpert and A.~B. Kahng, ``Recent directions in netlist partitioning: a
  survey,'' \emph{Integr. VLSI J.}, vol.~19, pp. 1--81, August 1995.

\bibitem{KDTree}
J.~L. Bentley, ``Multidimensional binary search trees used for associative
  searching,'' \emph{Commun. ACM}, vol.~18, pp. 509--517, September 1975.

\bibitem{FLOOR_TCAD96_SP}
H.~Murata, K.~Fujiyoshi, S.~Nakatake, and Y.~Kajitani, ``{VLSI} module
  placement based on rectangle-packing by the sequence-pair,'' \emph{IEEE
  Transactions on Computer-Aided Design of Integrated Circuits and Systems
  (TCAD)}, vol.~12, pp. 1518--1524, 1996.

\bibitem{Gurobi}
``{GUROBI},'' \url{http://www.gurobi.com/html/academic.html}.

\end{thebibliography}
